\newcommand{\up}{\uparrow}
\newcommand{\down}{\downarrow}
\newcommand{\nn}{\nonumber}

\documentclass{jpsj2}
\usepackage{psfrag}
\usepackage{latexsym}
\usepackage{graphicx}

\title{Nonlinear Optical Response of SDW Insulators}
\author{
S. Akbar \textsc{Jafari}$^{1,2}$\thanks{E-mail address: akbar@imr.tohoku.ac.jp}, 
Takami \textsc{Tohyama}$^{1}$
and Sadamichi \textsc{Maekawa}$^{1,3}$
}

\inst{$^{1}$ Institute for Materials Research,
Tohoku University, Sendai 980-8577, Japan \\
$^{2}$Department of Physics, Isfahan University of Technology, Isfahan
84156, Iran\\
$^{3}$CREST, Japan Science and Technology Agency, Kawaguchi
332-0012, Japan}

\abst{
We calculate the third order nonlinear optical response in the
Hubbard model within the spin density wave (SDW) mean field ansatz
in which the gap is due to onsite Coulomb repulsion. We obtain
closed-form analytical results in one dimension (1D) and two
dimension (2D), which show that nonlinear optical response in SDW
insulators in 2D is stronger than both 3D and 1D. We also
calculate the two photon absorption (TPA) arising from the stress
tensor term. We show that in the SDW, the contribution from stress
tensor term to the low-energy peak corresponding to two photon
absorption becomes identically zero if we consider the gauge
invariant current properly.
}
\kword{
Third harmonic generation, Two photon absorption, Gauge invariance,
SDW insulator
}

\recdate{\today}
\begin{document}
\maketitle

\section{Introduction} 
The realization of ultra fast networking through all-optical
switching in modern optical technology requires advanced optical
materials with large third-order nonlinear optical susceptibility
$\chi^{(3)}$(Ref. 1). Quasi one dimensional (1D)
$\pi$-conjugated polymers offer $\chi^{(3)}$ values of $10^{-12}$
to $10^{-7}$ e.s.u. (electronic system of units). The quasi 1D
Mott insulators such as Sr$_2$CuO$_3$, offer $\chi^{(3)}$ values
in the range $10^{-8}$ to $10^{-5}$
e.s.u.\cite{KishidaNature,KishidaPRL}.

   The canonical model for strongly correlated materials is the
standard Hubbard model. In the spin density wave (SDW) mean field
approximation, this Hamiltonian reduces to a simple quadratic form
that can be diagonalized exactly. Such an SDW Hamiltonian is
indeed relevant to several groups of organic
materials\cite{Gruner,Mazumdar}. The SDW approximation assumes
long range order. On the other hand, since such a model is
essentially non-interacting, there are no vertex corrections in
the current loops and the calculation of $\chi^{(3)}$ is greatly
simplified. This simplification allows us to calculate various
nonlinear optical processes, including those due to the stress
tensor.

   Although in one dimensional organic materials, where SDW states 
have been observed, people have already studied the optical
conductivity (linear response), nonlinear optical response in this
approximation which admits closed form expressions in 1D and 2D
has not been addressed yet. In this contribution we examine the
nonlinear optical response of SDW insulators in one, two and three
dimensions. Contrary to commonly accepted intuition that lower
dimensions is equivalent to larger optical responses, we find that
among SDW insulators, optical response in 2D larger that both one
and three dimensions.

   Among 1D, 2D and 3D systems with large on-site Coulomb
interaction, the 1D system has the largest optical nonlinearity
because of the decoupling of spin and charge degrees of
freedom\cite{Mizuno, HolonDoublon}. In contrast to this, among
SDW-ordered systems, the largest third order optical response
appears in 2D.

   Another purpose of this contribution is to clarify the importance
of gauge-invariant treatment in a simple model. From symmetry
point of view, some of optically allowed peaks (such as the case
of two photon absorption) may become zero, provided there is
charge conjugation symmetry\cite{Guo}. But, in SDW case, the mean
field factorization of the Hubbard model that leads to SDW
Hamiltonian, breaks this symmetry. However, when dealing with the
contribution arising from stress tensor terms, we find that the
gauge symmetry gives identically zero contribution to mid-gap peak
in TPA. This result implies that the mid-gap peak in TPA is solely
due to the four-current correlations. This observation is a
symmetry property, and as we will explicitly show, is independent
of dimension.

This paper is organized as follows: In section 2 we review the SDW
mean field treatment of the Hubbard model. In section 3 we discuss
the choice of gauge and method of calculation. In section 4 we
discuss the four-current contribution to the third harmonic
generation (THG) spectrum in 1D, 2D and 3D. In section 5 we
consider the effects of stress tensor terms that come through
quadratic couplings of gauge field to the electron system and also
through the dependence of gauge invariant current to the stress
tensor. Finally in section 6, we summarize the results.

\section{Model Hamiltonian}
   The SDW Hamiltonian is obtained from the Hubbard Hamiltonian
in the mean field approximation, where the gap is driven by
Coulomb repulsion. The Hubbard Hamiltonian is written as:
\begin{align}
   H &= \sum_{ks} \epsilon_k c^{\dagger}_{ks} c_{ks}
    + U \sum_j (n_{j\up}-n/2)(n_{j\down}-n/2),
\end{align}
where $c^{\dagger}_{ks}$ creates an electron with momentum $k$ and
spin $s=\up,\down$. The dimension of the lattice can be arbitrary,
but in this derivation let us focus on 1D case. We are interested
in half-filled case ($n=1$). Hereafter we will fix the scale of
energy by setting $2t_0=1$, where $t_0$ is the hoping amplitude.
In this paper we also use the system of units in which
$\hbar=c=e=a=1$, where $a$ is the lattice constant. The hopping
part of this Hamiltonian is characterized by the dispersion
$\epsilon_k = - \cos k$. The SDW mean field approximation amounts
to requiring
\begin{align}
   \langle n_{js} \rangle = \frac{n}{2} + s m e^{iQj},
\end{align}
where $n$ is the average particle density, and $s=\pm$ for $\up$
and $\down$, respectively, and $Q=\pi$. Ignoring fluctuations, and
dropping additive constant of energy, we obtain the quadratic SDW
Hamiltonian \cite{Fazekas}
\begin{align}
   H^{\rm SDW} &=& H_0 - Um \sum_j ~e^{iQj}~(n_{j\up}-n_{j\down}),
\end{align}
where $H_0$ is the tight-binding band part. This can be written in
a more compact form as
\begin{align}
   H^{\rm SDW} = \sum_{k}\sum_{s} \chi_{ks}^{\dagger} {\cal
   H}_{ks} \chi_{ks},
\end{align}
where
$\chi_{ks}^{\dagger}=(c_{ks}^{\dagger},c^{\dagger}_{k+Qs})=(c^{\dagger
c}_{ks},c^{\dagger v}_{ks})$ and
\begin{align}
   {\cal H}_{k\sigma} &=
   (\frac{\epsilon_k+\epsilon_{k+Q}}{2})
   +\frac{\epsilon_k-\epsilon_{k+Q}}{2} \sigma^z
   -Um\sigma\sigma^x. \nn
\end{align}
Here $k$ runs over the half BZ and $\sigma$'s are Pauli spin
matrices. If the perfect nesting property
$\epsilon_{k+Q}=-\epsilon_k$ holds, we have a much simpler
Hamiltonian
\begin{align}
   {\cal H}^{\rm SDW}_{ks} = \epsilon_k \sigma^z-s\Delta\sigma^x,
\end{align}
where the $k-$independent gap parameter $\Delta=Um$ is determined
by $U$.

   The unitary transformation $\psi_{k s}=U^{\dagger}\chi_{ks}$,
such that $U{\cal H}_kU^{\dagger}$ is diagonal, is given by
\begin{align}
   U^{\dagger} = \left( \begin{array}{cc}
      u_k    &   v_k \\
      -v_k^* &   u_k^*
   \end{array}\right),~~~~~~|u_k|^2+|v_k|^2=1,
\end{align}
with
\begin{align}
\begin{split}
   u_k = \frac{is}{\sqrt
   2}\sqrt{1+\frac{\epsilon_k}{\varepsilon_k}},\cr
   v_k = \frac{-i}{\sqrt
   2}\sqrt{1-\frac{\epsilon_k}{\varepsilon_k}},\cr
   \varepsilon_k = \sqrt{\Delta^2+\epsilon_k^2}.
\end{split}
\end{align}
We also need to note the relations:
\begin{align}
\begin{split}
   v_k^2-u_k^2 = |u_k|^2-|v_k|^2 =
   \frac{\epsilon_k}{\varepsilon_k},\\
   -v_k^*u_k-u_k^*v_k = 2u_kv_k = \frac{s\Delta}{\varepsilon_k},
\end{split}
\end{align}
giving
\begin{align}
\begin{split}
   U^{\dagger}\sigma^x U
   =-\frac{s\Delta}{\varepsilon_k}\sigma^z-\frac{\epsilon_k}{\varepsilon_k}\sigma^x,\cr
   U^{\dagger}\sigma^z U
   =+\frac{\epsilon_k}{\varepsilon_k}\sigma^z-\frac{s\Delta}{\varepsilon_k}\sigma^x,
   \label{UtransSDW.eqn}
\end{split}
\end{align}
that imply:
\begin{align}
   H^{\rm SDW} = \sum_{ks} \psi^{\dagger}_{ks} \varepsilon_k \sigma^z\psi_{ks}.
\end{align}

In 1D the density of states for this model is given by (Appendix B)
\begin{align}
   \rho(\omega) = \frac{|\omega|}{\sqrt{\omega^2-\Delta^2}}
   \frac{1}{\sqrt{w_1^2-\omega^2}}\label{rho1D.eqn},
\end{align}
where $w_1=\sqrt{1+\Delta^2}$.

   One can also consider other quadratic models with gap and
coherence factors, such as the $U=0$ limit of the ionic Hubbard 
model\cite{IonicHubbard}
\begin{small}
\begin{align}
   H^{\rm b} = -t_0 \sum_{\ell}\left( a^{\dagger}_{\ell}b_{\ell \pm 1}+
   b^{\dagger}_{\ell\pm1}a_{\ell}\right)
   +\sum_\ell \epsilon_A a^{\dagger}_{\ell}a_\ell
   +\sum_\ell \epsilon_B b^{\dagger}_{\ell}b_\ell.
\end{align}
\end{small}
where 'b' stands for band insulator. This Hamiltonian 
describes simple tight-binding insulator in which the gap is
due to difference $\epsilon_A=-\epsilon_B\equiv\Delta$ in site
energy, not the Coulomb correlation (in SDW $\Delta = Um$). This
model will differ from SDW insulator in the
$\epsilon_k\leftrightarrow\Delta$ replacements in their coherence
factors. Such a difference will affect the first order responses
dramatically, but it is easy to see that third order optical
response of this model is identical to SDW model.

\section{Choice of gauge }
The coupling of electromagnetic field  to matter can be described
in two gauges. One is a gauge in which vector potential is zero,
but the scalar potential is non-zero and given by $A_0 = -{\mathbf
E}{\bf .}{\mathbf r}$. In this gauge, the electric field of
radiation couples to the dipole moment of electrons, and hence one
needs the matrix elements of the position operator $\mathbf r$ to
calculate the response of the  matter to electromagnetic
perturbation. Working in this gauge is suitable for molecules and
small clusters. Because of the $\mathbf r$ operator, the
calculations in this gauge are sensitive to boundary conditions.
Moreover, for periodic boundary conditions, one has problem in
choosing the origin of the $\mathbf r$ coordinate. Therefore this
gauge is ill-defined in thermodynamic limit, and sensitive to
boundary conditions \cite{Gebhard1}.

  On the other hand, we have an alternative choice of working
in a gauge in which scalar potential is zero, while the vector
potential $\mathbf A$ is non zero. In this gauge the coupling
between the external field and electrons at the first order is via
the current operator: ${\mathbf\jmath}.{\mathbf A}$. In second
order the gauge field couples to electrons via the stress tensor
operator as ${\mathbf A}.{\mathbf \tau}.{\mathbf A}$, etc. Working
in this gauge is actually equivalent to Peierls substitution, and
hence we call it the Peierls gauge.

Without taking into account the effect of nonlinear couplings, and
nonlinear dependence of the gauge-invariant current on the vector
potential, the response function at  $n$'th order is given by:
\begin{align}
   \chi^{(n)}(\Omega;\omega_1,\ldots,\omega_n)=-
   \frac{n e^2 \delta_{n1}}{\epsilon_0 m \omega_1^2}\hat I +
   \frac{\chi^{(n)}_{jj}(\Omega;\omega_1,\ldots,\omega_n)} {\epsilon_0 i \Omega\omega_1\ldots\omega_n},
\end{align}
where $n-$current correlation function is given by
\begin{small}
\begin{align}
   & \chi^{(n)}_{jj}(\Omega;\omega_1,\ldots,\omega_n)=
   \frac{1}{n!}\left(\frac{i}{\hbar}\right)^n \frac{1}{V}
   \int dr_1\ldots dr_n \\ 
   & \int dt_1\ldots dt_n 
   \int dr dt ~ e^{i\Omega t-ik.r} \langle T_c 
   \jmath(r,t)\jmath(r_1,t_1)\ldots \jmath(r_n,t_n)
   \rangle.\nn
\end{align}
\end{small}
Here $T_c$ is the time-ordering operator along the Keldysh
path\cite{Wu} and $\jmath(r,t)$ is particle current operator.

   In general a $m-$point Keldysh Green's function has a tensor
structure due to two time branches. A unitary transformation to
the "physical" representation will give
$G_{\alpha\beta\gamma\ldots}$ where
$$
   \alpha,\beta,\gamma\in \{a,r \}
$$
Here $a$ stands for "advanced", while $r$ means "retarded". The
optical experiments measure the fully retarded components which
are given by nested commutators. On the other hand, the sum of
nested commutators is generated by $G_{arr\ldots},etc.$, where
only one of the indices is equal to $a$ and the rest are equal to
$r$ index \cite{YuLu}. As will be shown in appendix A, due to the
commutators and appropriate $\theta$ functions, this component of
Keldysh Green's function is very special, in the sense that we do
not really need to get into Keldysh machinery in order to
calculate the nonlinear optical response. As is shown in appendix
A, the fully retarded $G_{arr\ldots r}$ component can always be
calculated within the framework of equilibrium quantum field
theory. We first calculate the time ordered expectation values and
then analytically continue the result to ensure the correct
behavior of the poles. If we need other components of Keldysh
Green's functions (the fluctuation functions) that involve
anti-commutators, and are related to noise spectroscopy, Keldysh
formulation becomes inevitable.

  Therefore, in the case of optical response, we can forget the two
time branches, and denote the response function  at say, third
order by $\langle\jmath\jmath\jmath\jmath\rangle$, keeping in mind
that this is an ordinary {\em time ordered} expectation value and
should be analytically continued according to prescription of
appendix A.

  Another important point
is that the four-current scheme for calculation of the optical
response is reliable  far from zero frequency. Therefore, we do
not have to worry about the so called zero frequency divergence
(ZFD) in our calculations \cite{MXu}. To remove the unphysical ZFD
one has to calculate a few more correlation functions, but as far
as the behavior near resonance region is concerned, the
$\langle\jmath\jmath\jmath\jmath\rangle$ is sufficient. To
appreciate this point, let us look at the first order response in
two gauges ($\mu$ is the dipole operator and is proportional to
$\bf r$; $\jmath$ is current operator, and proportional to
$\dot{\bf r}$):
\begin{align}
   \chi_{\mu\mu}(\omega;\omega_1) &=
   \int_0^\beta d\tau~e^{i\omega\tau}
   \!\!\int_0^\beta d\tau_1~e^{i\omega_1\tau_1}
   \langle \mu(\tau)\mu(\tau_1)\rangle \\
   \chi_{\jmath\jmath}(\omega;\omega_1) &=
   \int_0^\beta d\tau~e^{i\omega\tau}
   \int_0^\beta d\tau_1~e^{i\omega_1\tau_1}
   \langle \jmath(\tau)\jmath(\tau_1)\rangle
\end{align}
where a factor of $2\pi\delta(\omega+\omega_\sigma)$, with
$\omega_\sigma=\omega_1$ does also multiply the right hand side.
We integrate by parts with respect to both time variables $\tau$
and $\tau_1$ to obtain:
\begin{align}
    & \chi_{\mu\mu}(\omega;\omega_1) =
    \int_0^\beta d\tau~e^{i\omega \tau}\int_0^\beta
    \langle \mu(\tau)\mu(\tau_1)\rangle
    d\left(\frac{e^{i\omega_1\tau_1}}{i\omega_1}\right)\nn\\
        &= \frac{1}{i\omega_1}\int_0^\beta d\tau~e^{i\omega \tau}
    \left[\langle\mu(\tau)\mu(\beta)\rangle
    -\langle\mu(\tau)\mu(0)\rangle\right]\nn\\
    &-\frac{1}{i\omega_1}\frac{1}{i\omega}
    \int_0^\beta d\tau_1~e^{i\omega_1\tau_1}
    \left[ \langle \mu(\beta)\jmath(\tau_1)\rangle
    -\langle \mu(0)\jmath(\tau_1)\rangle \right]\nn\\
    &+\frac{1}{i\omega_1}\frac{1}{i\omega}\chi_{\jmath\jmath}(\omega;\omega_1)
    \label{mumujj.eqn}
\end{align}
where we have used the fact that both $\omega_1$ and $\omega$ are
bosonic Matsubara frequency, $e^{i\omega_1\beta}=1$. Hence we see
that the relation $\chi_{\mu\mu}(\omega;\omega_1)=
\frac{1}{i\omega_1}\frac{1}{i\omega}\chi_{\jmath\jmath}(\omega;\omega_1)$,
often used to relate the response in two gauges {\em is not quite
correct}. Inclusion of the boundary terms (first two lines of the
above equation) compensates the frequency denominators appearing
in $\langle\jmath\jmath\rangle$ scheme. Similar consideration
applies to higher order correlations such as
$\langle\jmath\jmath\jmath\jmath\rangle$.

   If we keep nonlinear coupling between the gauge field and electrons that
comes through the stress tensor operator $\mathbf \tau$, and also
dependence of gauge invariant current to powers of gauge field
$A$, a simple perturbation theory gives expressions of the type
$\langle\jmath\jmath\tau\rangle$ that again appear in fully
retarded combination of nested commutators. The simplicity of SDW
and our band model allows us to consider these terms as well.

\section{Four-current response}
  The particle current operator is given by
\begin{align}
   \jmath &= it_0 \sum_\ell \left( c^{\dagger}_{\ell+1}c_{\ell}
   -c^{\dagger}_{\ell}c_{\ell+1}\right)
\end{align}
The momentum space representation of this operator for the band
and SWD insulators is given by:
\begin{align}
   \jmath &= \sum_{ks}\chi^{\dagger}_{ks} \gamma_k \sigma^z \chi_{ks}
   ,~~~~~\gamma_k =2t_0\sin k
\end{align}
which in terms of new fermions is
\begin{align}
   \jmath^{\rm SDW} &=\sum_{ks} \psi^{\dagger}_{ks} \left(\frac{\gamma_k\epsilon_k}{\varepsilon_k}\sigma^z
   -\frac{\gamma_k\Delta}{\varepsilon_k}\sigma^x \right)\psi_{ks}
\end{align}
The coefficient of  $\sigma^z$ in the above expression describes
{\em intra}-band transitions, while the coefficient of $\sigma^x$
causes {\em inter}-band transitions. Let us define the
corresponding coefficients by
\begin{align}
    g \varepsilon_k = \sin k\cos k,~~~ h\varepsilon_k =s\Delta\sin k,
\end{align}
The THG susceptibility corresponding to photon frequency $\nu$ (or
$i\nu$ in imaginary time) is given by

A few comments are in order: The trace operator arises from
closing the current loop, and enormously simplifies the
calculations. This trace is taken with respect to the indices of
Pauli matrix. But since Pauli matrices along with unit matrix have
a closed algebra, and that only the unit operator survives the
trace, the sums of 16 terms in the case of first order response,
and of 256 terms in the case of third order response are
considerably simplified.

We need one more physical considerations to further simplify the
calculations. At half filled situation of interest to us, the
intra-band terms at zero temperature do not contribute to optical
absorption. In calculating general expectation value of say
$\langle ABCD\rangle$, where $A,B,C,D$ can be any operators
contributing to the coupling of light with matter, we are
interested in dominant processes in which terms containing $g$
(intra-band matrix element) in the rightmost and leftmost
operators $A$ and $D$ do not contribute. Similarly in $B$ and $C$
operators the $h$ term should be dropped.

   To calculate the time ordered product within the framework of
equilibrium quantum field theory, we use the Matsubara technique.
The summation over fermion loop frequency $i\omega_n$ can be done
with standard contour integration techniques. Also, if we ignore
the momentum from the incident light, the momentum $k$ running in
the current loop must be the same for all fermion propagators.
After contracting various fermion spinors to get the appropriate
Greens' functions we obtain
\begin{small}
\begin{align}
   &\frac{1}{\nu^4}\langle \jmath\jmath\jmath\jmath\rangle = 
   \sum_{k} \frac{1}{\nu^4 \varepsilon_k^3}\times\\
   &\frac{8(4\varepsilon_k^2+\nu^2)\Delta^2\cos(k)^2\sin (k)^4}
  {(2\varepsilon_k-3\nu )(2\varepsilon_k+3\nu)
  ( 2 \varepsilon_k -2\nu)( 2 \varepsilon_k+2\nu )
  ( 2 \varepsilon_k -\nu)( 2 \varepsilon_k+\nu )}.\nn
  \label{jjjj2.eqn}
\end{align}
\end{small}
The analytic continuation $\nu\to\nu+i0^+$ is implicitly
understood.

  It is also important to note that, as far as the behavior of
response functions near the resonance is concerned (that is away
from $\nu=0$), four-current response functions, $\langle
\jmath\jmath\jmath\jmath\rangle$ give the same qualitative
features as that of four-dipole response functions,
$\langle\mu\mu\mu\mu\rangle$, where $\mu$ is the dipole moment
operator. To appreciate this point, let us go back to the
expression of the current operator for the SDW insulator. The
inter-band matrix element for the current operator is $ \jmath_k =
s\Delta\sin k/\varepsilon_k $, that via the equation of motion for
the dipole moment operator $\mu$, would imply that the dipole
matrix elements are:
\begin{align}
    \mu_k = s\Delta\sin k/2\varepsilon_k^2.
\end{align}
Similar procedure that lead to equation (\ref{jjjj2.eqn}) gives:
\begin{small}
\begin{align}
    &\langle \mu\mu\mu\mu\rangle = 
    \sum_{k} \frac{1}{2\varepsilon_k^7}\times\\
    &\frac{(4\varepsilon_k^2+\nu^2)\Delta^2\cos(k)^2 \sin (k)^4}
  {(2\varepsilon_k-3\nu )(2\varepsilon_k+3\nu)
  ( 2 \varepsilon_k -2\nu)( 2 \varepsilon_k+2\nu )
  ( 2 \varepsilon_k -\nu)( 2 \varepsilon_k+\nu )},\nn
  \label{mmmm2.eqn}
\end{align}
\end{small}
Now we can see that up to a numerical factor of the order of
unity, the "near resonance" behavior of both expression is the
same. Because after analytic continuation, the imaginary parts
give delta functions peaking near $\sim \varepsilon_k$. Therefore
$\nu^4\varepsilon_k^3$ in the denominator of
$\langle\jmath\jmath\jmath\jmath\rangle $ is of the same order of
magnitude as $\varepsilon_k^7$ in $\langle \mu\mu\mu\mu\rangle $
expression. Note that four-current formula makes sense near the
resonance conditions only. It gives the unphysical $\nu^{-4}$
divergence in the static limit $\nu\to 0$. Fixing this problem as
discussed in equation (\ref{mumujj.eqn}), requires calculation of
a few more expectation values, but we are not interested in this
limit here. Since we are dealing with a gapped situation in which
the frequencies of interest are of the order of gap and zero
frequency is avoided.

  The imaginary part of THG susceptibility can now be written as
\begin{align}
   \Im\chi^{\rm THG}(\nu)=\frac{27}{2}f_D\left(\frac{3\nu}{2}\right)
   -8f_D(\nu)+\frac{1}{2}f_D\left(\frac{\nu}{2}\right)\cr
   f_D(\nu) = \sum_{k}\frac{\pi\Delta^2\cos^2k~\sin^4k}
   {24\varepsilon^6_k}\delta(\varepsilon_k-\nu),
\end{align}
where $D$ is the dimension of space. In 1D this integral can be
evaluated as follows:
\begin{align}
   f_1(\nu) &= \int d\varepsilon \rho(\varepsilon)
   \frac{\pi\Delta^2(\varepsilon^2-\Delta^2)(w_1^2-\varepsilon^2)^2}
   {12\varepsilon^6}\delta(\varepsilon-\nu)\nn\\
   &= \frac{\pi\Delta^2}{24|\nu|^5}
   \left(w_1^2-\nu^2\right)^{3/2}\left(\nu^2-\Delta^2\right)^{1/2}.
\end{align}
Using the Kramers-Kronig relation
\begin{align}
   \Re\chi(\omega)=\frac{1}{\pi}{\cal P}\int
   d\nu\frac{\Im\chi(\nu)}{\nu-\omega},
\end{align}
one can obtain a closed form expression for the real part of the
above retarded susceptibility in terms of Elliptic functions of
various kind.
\begin{figure}[t]
   \centering
   \psfrag{chi}{$\Im\chi^{\rm THG}(\nu)$}
   \psfrag{nu}{$\nu$}
   \psfrag{10 thg-1D}{$10\times \Im\chi^{\rm THG}_{1}(\nu)$}
   \psfrag{thg-2D}{$\Im\chi^{\rm THG}_{2}(\nu)$}
   \psfrag{100 thg-3D}{$100\times \Im\chi^{\rm THG}_{3}(\nu)$}
   \includegraphics[height=0.33\textwidth,angle=0]{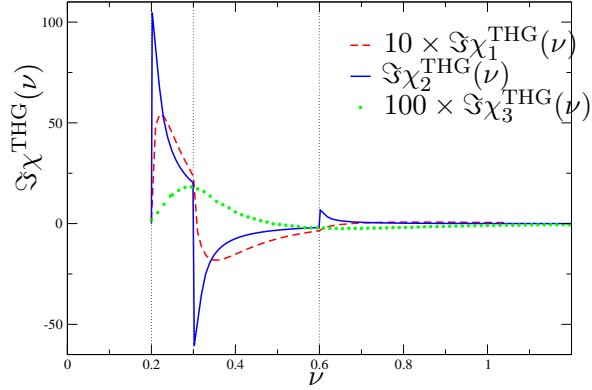}
   \caption {Plot of imaginary part of $\chi^{\rm THG}_D(\nu)$ vs.
   $\nu$ in the SDW model in $D=1,2,3$ dimensions.
   The unit of energy is $2t_0=1$, and in all calculations we have
   assumed $\Delta=0.3$. The gap is given by $E_g=2\Delta=0.6$.
   We see that the response in D=2 is stronger
   than D=1,3. Values of D=1,3 are magnified by factors of $10$ and $100$,
   respectively, for eye assistance.}
   \label{chi3-3D.fig}
\end{figure}

   In 2D still we can obtain closed form results bye transforming
to tight binding coordinates (appendix B)
\begin{align}
   f_2(\nu/2) &= \frac{\Delta^2}{6\pi\nu\lambda}
   \int_{\Delta}^{w_2}\!\! d\varepsilon
   \int_0^{\pi/4}\!\!d\xi\frac{\delta(\nu/2-\varepsilon)}
   {\sqrt{1-\beta\cos^2(2\xi)}}\nn\\
   &=\frac{\Delta^2}{90\pi\nu\lambda}
   E\left(\sqrt\beta\right)\left[548\beta^2-1258\beta-708\right]\nn\\
   &+\frac{\Delta^2}{90\pi\nu}\frac{\lambda}{4}
   K\left(\sqrt\beta\right)\left[(60\beta^2-226\beta+177)\right].
\end{align}
In this expression $\lambda^2=\nu^2/4-\Delta^2$. Here $E$ and $K$
are elliptic functions of second and first kinds,
respectively\cite{Abramowitz}. The second term near the band edge
behaves like $\sqrt{\nu/2-\Delta}\log(\nu/2-\Delta)$ which remains
finite, while the first term give a divergent contribution of the
form $f_2(\nu/2)\sim 1/\sqrt{\nu/2-\Delta}$. The main THG peak is
due to $f_2(3\nu/2)\sim 1/\sqrt{3\nu/2-\Delta}$ which occurs at
$\nu=2\Delta/3$, with $2\Delta$ being the excitation gap.

   In three dimensions $f_3(\nu)$ and hence $\Im\chi^{\rm THG}(\nu)$
can be calculated numerically. Figure \ref{chi3-3D.fig} shows
$\Im\chi^{\rm THG}_D(\nu)$ for $D=1,2,3$ and the gap parameter
$\Delta=0.3$. The 3D result is magnified $100$ times (dotted
line), and the 1D result (dashed line) is magnified $10$ times. In
the SDW system, inverse square root divergence in 2D has no
counterpart in 1D and 3D. To understand this, notice that, one can
always replace a $k$ integration with a energy integration
weighted by DOS, and so such an enhancement in 2D compared to 1D
and 3D can be traced back to the nature of singularities of DOS
(appendix B). This figure demonstrates that if we have a insulator
in which gap is due to SDW type of order, the phase space effect
(DOS) along with SDW coherence factors ($u_k$, $v_k$) generate
larger nonlinear response in 2D compared to 1D and 3D.

\section{Stress tensor terms}
The simplicity of SDW model, along with possibility of obtaining
closed form results in one and two dimensions allows us to
investigate the role of stress tensor terms in nonlinear optical
processes. The first place that stress tensor appears is via the
coupling of external field to matter at second order which is
$\tau.{\bf A}.\tau$. Then the gauge invariant current (particle
current plus a correction from external field) involves the stress
tensor itself, and to first order is given by\cite{Gebhard1}:
\begin{align}
    {\cal J}^m = \jmath^m-\tau^{mn} A_n
    \label{gauge-invariant-current.eqn},
\end{align}

   Therefore the third order response of gauge invariant current
involves the response of both {\em particle-current} operator
($\jmath^m$ above) and also the response of stress tensor operator
($\tau^{mn}$ above). The third order response due to
particle-current is
\begin{align}
   \langle\jmath\tau\jmath\rangle +
   \langle\tau\jmath\jmath\rangle
   \label{particle-j-response.eqn}.
\end{align}
while the third order response due to $\tau$ operator is
\begin{align}
   \langle\jmath\jmath\tau\rangle
\end{align}
These expressions are of the type given in (\ref{chiABQ.eqn}). The
first two corresponds to $A,Q\to\jmath,~B\to\tau$, and
$A\to\tau,~B,Q\to\jmath$, respectively, while the third one
corresponds to $A,B\to\jmath,~Q\to\tau$.

  To obtain the appropriate matrix elements, let us begin by
writing the stress tensor in 1D as
\begin{align}
   \tau = -t_0\sum_{ns}\left(c^{\dagger}_{ns}c_{n+1s}+c^{\dagger}_{n+1s}c_{ns}\right),
\end{align}
or equivalently
$\tau=\sum_{ks}\chi^{\dagger}_{ks}\epsilon_k\sigma^z\chi_{ks}$,
that eventually becomes
\begin{align}
   \tau^{\rm SDW} &= \sum_{ks} \psi^{\dagger}_{ks}\left(
   \frac{\epsilon_k^2}{\varepsilon_k}\sigma^z-
   \frac{s\Delta\epsilon_k}{\varepsilon_k}\sigma^x
   \right)\psi_{ks}.
\end{align}

The matrix elements we need are given in table
\ref{matrix-elements.tab},
\begin{table}[htb]
\begin{center}
\begin{tabular}{|l|l|l|}
       \hline
       &  $\jmath^{\rm inter}\jmath^{\rm intra}\tau^{\rm inter}$ &
       $\jmath^{\rm inter}\tau^{\rm intra}\jmath^{\rm inter}$ \\
       \hline
       SDW  & ${\Delta^2\sin^2k\cos^2k}/{\varepsilon^3_k}$  &
       $s{\Delta^2\sin^2k\cos^2k}/{\varepsilon^3_k}$ \\
       \hline
\end{tabular}
\caption{Inter band and intra band matrix elements in the SDW
model.} \label{matrix-elements.tab}
\end{center}
\end{table}
where $s$ is $\pm 1$ for $\up$ and $\down$ spins, respectively.
Since the matrix elements involve both $\sin$ and $\cos$, in 1D
the transitions at zone center and zone boundary are suppressed,
and hence there is no divergence. But in 2D again SDW response
will be divergent.

  Let us look at the two photon absorption (TPA) contribution
arising from $\tau$ operator carefully. In ordinary case of
$\langle \jmath\jmath\jmath\jmath\rangle$, the TPA corresponds to
$\omega_1=\omega_2=-\omega_3=\nu$. In the case of stress tensor
TPA corresponds to $\omega_1=\omega_2=\nu$. If we denote the
denominators of the first, second and third lines of the
expression (\ref{chiABQ.eqn}) by $\ell_1,\ell_2,\ell_3$,
respectively, after decomposing to partial fractions we have
\begin{align}
   \ell_{1,2} &= \frac{1}{2\varepsilon_k(\nu+i\eta-\varepsilon_k)}
              -\frac{1}{2\varepsilon_k(\nu+i\eta-2\varepsilon_k)},\\
   \ell_3 &= \frac{1}{2\varepsilon_k(\nu+i\eta-2\varepsilon_k)},
\end{align}
where a $(\nu+i\eta\to-\nu-i\eta)$ term also is present in the
time ordered correlation function. Putting in the matrix elements,
the TPA susceptibility becomes
\begin{align}
   &\left(\jmath^{\rm inter} \right)^2\tau^{\rm intra}
   \left[\ell_1+\ell_2-\ell_3 \right]\\
   &+\jmath^{\rm inter}\jmath^{\rm intra}\tau^{\rm inter}
   \left[(\ell_2+\ell_3)+(\ell_3+\ell_1)-(\ell_1+\ell_2)\right],\nn
\end{align}
where we have not simplified the second line deliberately to
emphasize the role of gauge invariant current. Obviously what we
measure is the gauge invariant current. The minus sign in third
terms of the each line in the above equation comes from the minus
sign in equation (\ref{gauge-invariant-current.eqn}), which is
essential for the cancellation that takes place in the second line
of the above equation, leaving a contribution proportional to
 $\propto \ell_3$.

   Now let us concentrate on the first line of this equation.
Here the matrix element have opposite signs for up and down spins
(which ultimately comes from the coherence factors of the SDW
state). Therefore their contributions identically cancel each
other, and we are left only with a term coming from denominator of
$\ell_3$ only, which peaks at the gap $2\Delta$. In 1D as we
mentioned, the presence of $\sin$ and $\cos$ suppresses the
transitions at band edge and zone center. So let us calculate this
term in 2D in the SDW insulator:
\begin{small}
\begin{align}
\begin{split}
   &\langle (\jmath_x\jmath_x\tau_{xx})\rangle^{\rm SDW} =
   -\frac{\pi\Delta^2}{4}\sum_{k_x,k_y} \delta(\nu/2-\varepsilon_k)
   \frac{\sin^2k_x\cos^2k_x}{\varepsilon_k^4}\\
   &=-\frac{\pi\Delta^2}{4}\int\!\!
       \int\frac{dxdy}{2\pi^2}\frac{\sin^2x\cos^2x}{\varepsilon^4}
   \delta(\nu/2-\varepsilon)\\
   &=-\frac{\Delta^2}{8\pi}\int\!\!\int\frac{d\lambda d\xi}{J}
      \frac{\sin^2x\cos^2x}{\varepsilon^4}
   {\delta(\nu/2-\varepsilon)}\\
   &=-4\Delta^2\frac{\lambda(12\beta-23)K(\sqrt{\beta})}{3\pi\nu^4}
   -16\frac{(23-22\beta)E(\sqrt\beta)}{3\pi\lambda\nu^4},
\end{split}
\end{align}
\end{small}
where in the last line again delta function picks up the value of
$\lambda=\sqrt{\nu^2/4-\Delta^2}$, $\beta=1-\lambda^2/4$ as in the
appendix B. The logarithmic divergence of the first term gets
suppressed by $\lambda\sim\sqrt{\nu-2\Delta}$ factor, but the
second term still shows inverse square root divergence in 2D.
Therefore in the case of stress tensor terms too, the 2D SDW
systems offer larger response than 1D.

   We would also like to emphasize that, the requirement of
gauge invariance cancels the singularity at $\Delta=E_g/2$, in
$\tau$'s contribution and leaves us with a square root singularity
in $2\Delta=E_g$ in 2D SDW systems. In other words, in SDW
insulator the peak at $E_g/2$ may solely be due to four-current
correlations and the contribution from stress tensor to this peak
is identically zero.

\section{Summary and conclusions}
In conclusion, we have calculated the nonlinear optical responses
in the SDW insulator in which the gap is due to on-site Coulomb
repulsion. The linear response of SDW model has the characteristic
inverse square root divergence in 1D which due to larger amount of
nesting is further enhanced by a logarithmic factor in 2D and is
entirely suppressed in 3D. The THG spectrum of SDW model has no
divergence in 1D, but diverges as inverse square root in 2D and
becomes finite again in 3D. Therefore the optical responses
(linear and nonlinear) of SDW insulators is maximal in 2D as a
function of dimensionality.

  The model calculations presented in this work suggests that
{\em nesting} as a possible mechanism of nonlinearity enhancement
which works best in 2D, contrary to common intuition that lower
spatial dimensions are better for nonlinear optical materials.
This mechanism does not work in 1D. It was found that in 1D such a
enhancement can arise from the spin-charge
separation\cite{Mizuno,HolonDoublon}.

  The simplicity of the quadratic model treated in this investigation
allowed us to exactly calculate the contributions of stress tensor
in nonlinear optical response. We showed that (i) in TPA
measurements in the SDW systems, the structure at the mid-gap
($E_g/2$) has essentially no contribution from the stress tensor
term, and (ii) when these contributions are non-zero, have
comparable effect to that of usually considered four-current
terms. Stress term has even parity and in nonlinear processes can
lead to dipole-forbidden transition. This hints to the importance
of gauge invariant treatment of currents in nonlinear optics,
which is usually neglected in the literature.

\section{Acknowledgments}
This work was supported by Grant-in-Aid for Scientific Research,
MEXT of Japan, CREST, and NAREGI. S.A.J. was supported by JSPS
fellowship P04310.

\appendix
\section{Analytic continuation}
In nonlinear response theory, we need to calculate fully retarded
expectation value of nested commutators. For example in the case
of nonlinear dielectric response of an electron gas to a fast
moving ion, the nested commutator of density operators at
different times appears \cite{Bergara}. In the general theory of
nonlinear response, we might have the fully retarded combination
of nested commutators of arbitrary operators. These operators are
determined by nature of coupling (linear, quadratic, etc.) between
the system and the external perturbation, and also the observable
being studied. For example the second order coupling of the
electromagnetic field to matter via the stress tensor operator
$\tau$, leads to a fully retarded  current response of the form
$\langle [\jmath,[\jmath,\tau]] \rangle$. The general framework to
study this type of expectation values is the non-equilibrium
quantum field theory. However, within the standard formulation of
quantum field theory, with appropriate analytic continuation, one
can obtain these kind of expectation values that correspond to
$G_{arr\ldots r}$ component in Keldysh Green's function language.
In optical measurements always this kind of expectation values
appear. In noise spectroscopies the other components of Keldysh
Green's function appear that can not be treated as straightforward
as $G_{arr\ldots r}$ component, and use of Keldish formulation
becomes necessary.

  For the problem in which external perturbation couples linearly
to the system, through operator $A_j$, and quadratically through
$B_{kl}$, where we are interested in variations of quantity $Q_i$,
one can write the retarded response at third order as:
\begin{small}
\begin{align}
   &\phi_{ijkl}^R(t;t_1,t_2) = +\theta(t-t_2)\theta(t_2-t_1)
   \langle [A_j(t_1),[B_{kl}(t_2),Q_i(t)]] \rangle\nn\\
   &+\theta(t-t_1)\theta(t_1-t_2)
   \langle [B_{kl}(t_2),[A_j(t_1),Q_i(t)]]\rangle\nn\\
   &+(j\to k\to l\to j) + (j\to k\to l\to j)^2.
\end{align}
\end{small}

  On the other hand, the time-ordered product of these three
operators is given by:
\begin{align}
\begin{split}
    &\phi_{ijkl}^T(t;t_1,t_2) = \langle T A_j(t_1)B_{kl}(t_2)Q_i(t)\rangle=\\
    &+\theta(t_1-t_2)\theta(t_2-t) \langle  A_j(t_1)B_{kl}(t_2)Q_i(t) \rangle\\
    &+\theta(t-t_2)\theta(t_2-t_1) \langle Q_i(t)B_{kl}(t_2)A_j(t_1)\rangle \\
    &+\theta(t_2-t_1)\theta(t_1-t) \langle B_{kl}(t_2)A_j(t_1)Q_i(t)\rangle \\
    &+\theta(t-t_1)\theta(t_1-t_2) \langle Q_i(t)A_j(t_1)B_{kl}(t_2)\rangle \\
    &+\theta(t_1-t)\theta(t-t_2) \langle A_j(t_1)Q_i(t)B_{kl}(t_2)\rangle \\
    &+\theta(t_2-t)\theta(t-t_1) \langle B_{kl}(t_2)Q_i(t)A_j(t_1)\rangle\\
    &+permutations.
\end{split}
\end{align}
The signs are all positive, since operators $A,B,Q$ are quadratic
in fermion operators.

   Now let us expand the commutators in definition of retarded
expectation value to obtain
\begin{align}
   &\phi_{ijkl}^R(t;t_1,t_2) =\nn\\
   &+\theta(t-t_2)\theta(t_2-t_1)\langle A_j(t_1)B_{kl}(t_2)Q_i(t) \rangle\nn\\
   &-\theta(t-t_2)\theta(t_2-t_1)\langle A_j(t_1)Q_i(t)B_{kl}(t_2)\rangle\nn\\
   &-\theta(t-t_2)\theta(t_2-t_1)\langle B_{kl}(t_2)Q_i(t)A_j(t_1)\rangle\nn\\
   &+\theta(t-t_2)\theta(t_2-t_1)\langle Q_i(t)B_{kl}(t_2)A_j(t_1)\rangle\nn\\
   &+\theta(t-t_1)\theta(t_1-t_2)\langle B_{kl}(t_2)A_j(t_1)Q_i(t)\rangle\nn\\
   &-\theta(t-t_1)\theta(t_1-t_2)\langle B_{kl}(t_2)Q_i(t)A_j(t_1)\rangle\nn\\
   &-\theta(t-t_1)\theta(t_1-t_2)\langle A_j(t_1)Q_i(t)B_{kl}(t_2)\rangle\nn\\
   &+\theta(t-t_1)\theta(t_1-t_2)\langle Q_i(t)A_j(t_1)B_{kl}(t_2)\rangle.\nn
\end{align}
Now using the identity
\begin{small}
\begin{align}
   \theta(t-t_2)\theta(t_2-t_1)+\theta(t-t_1)\theta(t_1-t_2)=\theta(t-t_2)\theta(t-t_1),
\end{align}
\end{small}
the sum of second and seventh terms simplifies to
\begin{align}
    -\theta(t-t_2)\theta(t-t_1)\langle AQB \rangle,
\end{align}
while the sum of third and sixth terms becomes
\begin{align}
    -\theta(t-t_2)\theta(t-t_1)\langle BQA \rangle,
\end{align}
so that we obtain
\begin{align}
\begin{split}
   &\phi_{ijkl}^R(t;t_1,t_2) =\\
   &+\theta(t-t_2)\theta(t_2-t_1)\langle A_j(t_1)B_{kl}(t_2)Q_i(t)\rangle\\
   &+\theta(t-t_2)\theta(t_2-t_1)\langle Q_i(t)B_{kl}(t_2)A_j(t_1)\rangle\\
   &+\theta(t-t_1)\theta(t_1-t_2)\langle B_{kl}(t_2)A_j(t_1)Q_i(t)\rangle\\
   &+\theta(t-t_1)\theta(t_1-t_2)\langle Q_i(t)A_j(t_1)B_{kl}(t_2)\rangle\\
   &-\theta(t-t_1)\theta(t-t_2)\langle A_j(t_1)Q_i(t)B_{kl}(t_2)\rangle\\
   &-\theta(t-t_1)\theta(t-t_2)\langle B_{kl}(t_2)Q_i(t)A_j(t_1)\rangle.
\end{split}
\end{align}
Now apart from the $\theta$ function that determines the
analytical structure, the time ordered and retarded nested
commutators have similar structures. Therefore one can obtain the
expectation value of fully retarded nested commutator by
appropriate analytic continuation of the corresponding time
ordered one. So, let us obtain the spectral representations of the
retarded and time ordered expectation values and compare them.

Our conventions for Fourier transforms are
\begin{small}
\begin{align}
   f(t) = \int_{-\infty}^{+\infty}\frac{d\omega}{2\pi}
   e^{-i\omega t}\tilde f(\omega) \Leftrightarrow
   \tilde f(\omega) = \int_{-\infty}^{+\infty}dt e^{+i\omega t} f(t),\cr
   \chi(\omega;\omega_1,\omega_2) = \int\!\frac{d\omega}{2\pi}
   \int\!\frac{d\omega_2}{2\pi}
   \int\!\frac{d\omega_1}{2\pi}
   e^{i\omega t}e^{i\omega_1 t_1}e^{i\omega_2 t_2}\chi(t;t_1,t_2).
\end{align}
\end{small}
Note that the operator $B(t_2)$ acting at time $t_2$ couples the
second power of the external perturbation to the system.
Therefore, in principle it could involve two frequencies
$\omega_2,\omega_3$. But since the external fields are supposed to
act at the same time, we always have the combination
$\omega_2+\omega_3$. Therefore here we have dropped the $\omega_3$
in our calculations. Using the representation
\begin{align}
   \theta(t) = i\int_{-\infty}^{+\infty}\frac{d\omega}{2\pi}\frac{e^{-i\omega t}}{\omega+i\eta},
\end{align}
for the step function, after some algebra we get
\begin{align}
\begin{split}
   &\chi_T(\omega;\omega_1,\omega_2) = 2\pi \delta(\omega+\omega_{\sigma})\sum_{a,b}\\
   &-\frac{A_j^{0a}B_{kl}^{ab}Q_i^{b0}}
         {(\omega_{\sigma}+E_{b0}-i\eta)(\omega_1-E_{0a}-i\eta)}\\
   &-\frac{Q_i^{0a}B_{kl}^{ab}A_j^{b0}}
         {(\omega_{\sigma}+E_{0a}+i\eta)(\omega_1-E_{b0}+i\eta)}\\
   &-\frac{B_{kl}^{0a}A_j^{ab}Q_i^{b0}}
         {(\omega_{\sigma}+E_{b0}-i\eta)(\omega_2-E_{0a}-i\eta)}\\
   &-\frac{Q_i^{0a}A_j^{ab}B_{kl}^{b0}}
         {(\omega_{\sigma}+E_{0a}+i\eta)(\omega_2-E_{b0}+i\eta)}\\
   &+\frac{A_j^{0a}Q_i^{ab}B_{kl}^{b0}}
         {(\omega_2-E_{b0}+i\eta)(\omega_1-E_{0a}-i\eta)}\\
   &+\frac{B_{kl}^{0a}Q_i^{ab}A_j^{b0}}
         {(\omega_2-E_{0a}-i\eta)(\omega_1-E_{b0}+i\eta)},
\end{split}
\end{align}
while
\begin{align}
\begin{split}
   &\chi_R(\omega;\omega_1,\omega_2) = 2\pi \delta(\omega+\omega_{\sigma})\sum_{a,b}\\
   &-\frac{A_j^{0a}B_{kl}^{ab}Q_i^{b0}}
         {(\omega_{\sigma}+E_{b0}+i\eta)(\omega_1-E_{0a}+i\eta)}\\
   &-\frac{Q_i^{0a}B_{kl}^{ab}A_j^{b0}}
         {(\omega_{\sigma}+E_{0a}+i\eta)(\omega_1-E_{b0}+i\eta)}\\
   &-\frac{B_{kl}^{0a}A_j^{ab}Q_i^{b0}}
         {(\omega_{\sigma}+E_{b0}+i\eta)(\omega_2-E_{0a}+i\eta)}\\
   &-\frac{Q_i^{0a}A_j^{ab}B_{kl}^{b0}}
         {(\omega_{\sigma}+E_{0a}+i\eta)(\omega_2-E_{b0}+i\eta)}\\
   &+\frac{A_j^{0a}Q_i^{ab}B_{kl}^{b0}}
         {(\omega_2-E_{b0}+i\eta)(\omega_1-E_{0a}+i\eta)}\\
   &+\frac{B_{kl}^{0a}Q_i^{ab}A_j^{b0}}
         {(\omega_2-E_{0a}+i\eta)(\omega_1-E_{b0}+i\eta)},
         \label{chiABQ.eqn}
\end{split}
\end{align}
where we have defined $\omega_{\sigma}=\omega_2+\omega_1$. We see
the exact parallelism between the time ordered and fully retarded
expectation values. The important difference is due to the nature
of step functions that give rise to $\omega_n+i\eta$
($n=1,2,\sigma$) structure in the retarded function. This is what
was expected from causality imposed by appropriate $\theta$
functions in fully retarded one. However, as a consequence of
having $\theta$ functions along with commutators, equal time
contractions in diagrammatic perturbation theory do not
contribute, and should be excluded.

 We can also write the above result in a more compact form if
we note that the frequencies are associated with operators as
$(A,\omega_1),(B,\omega_2),(Q,-\omega_\sigma)$. If we ignore the
$i\eta$ factor, the above result can be written as
\begin{small}
\begin{align}
   \chi_R(\omega;\omega_1,\omega_2) = 2\pi \delta(\omega+\omega_{\sigma})
   \sum_{a,b}\sum_{\cal P}\frac{A_j^{0a}B_{kl}^{ab}Q_i^{b0}}
         {(-\omega_{\sigma}+E_{0b})(\omega_1-E_{0a})},
\end{align}
\end{small}
where ${\cal P}$ stands for all different permutations of
$(A,\omega_1),(B,\omega_2),(Q,-\omega_\sigma)$. At the end we must
remember to use the appropriate $i\eta$ factors to ensure
$\omega_n+i\eta$ structure. One can easily see that this
prescription gives the correct spectral representation in case of
two-current correlation with $A=B=\jmath$:
\begin{align*}
   \chi_R(\omega;\omega_1)=2\pi\delta(\omega+\omega_1)
   \sum_a\frac{|\langle 0|\jmath |a\rangle|^2}{\omega_1-(E_a-E_0)+i0^+}.
\end{align*}

   We can derive a similar prescription for higher order
correlation functions of fully retarded nature in a
straightforward way:
\begin{align}
\begin{split}
   &\chi_R(\omega;\omega_1,\omega_2,\omega_3) =
      2\pi \delta(\omega+\omega_{\sigma}) \sum_{a,b,c}\cr
   &\sum_{\cal P}\frac{A^{0a}B^{ab}C^{bc}Q^{c0}}
         {(-\omega_{\sigma}+E_{0c})(\omega_2+\omega_1-E_{0b})(\omega_1-E_{0a})}.
\end{split}
\end{align}

\section{Tight binding coordinates}
In this appendix, we denote the $k_x$ and $k_y$ coordinates in the
reciprocal space by $x,y$ for convenience. Since constant energy
surfaces $\cos x+\cos y$ appear very frequently in calculations
related to 2D tight binding systems, it is useful to define a {\em
natural} orthogonal transformation, $(x,y)\to (\lambda,\xi)$, so
that constant coordinate surfaces correspond to constant energy
surfaces. The first coordinate obviously must be
\begin{align}
   \lambda = \cos x +\cos y.
   \label{lambdacoord.eqn}
\end{align}
In order to guess an appropriate form for the second coordinate
$\xi$, we require the constant $\xi$ surfaces to be orthogonal to
the constant $\lambda$ surfaces, that is
\begin{align}
   \vec v_\xi \propto \nabla\lambda = -\sin x~ \hat e_x -\sin y~\hat e_y,
\end{align}
where $\vec v_\xi$ is a 'velocity' tangent to the constant $\xi$
surface. This equation implies that
\begin{align}
   \frac{dx}{dt}=-\sin x,~\frac{dy}{dt}=-\sin y,
\end{align}
the division of which gives
\begin{align}
   \frac{dx}{\sin x} = \frac{dy}{\sin y}\Rightarrow
   d\ln \left(\tan (x/2)\right) = d\ln\left(\tan (y/2)\right).
\end{align}
Integrating the above equation gives $\tan(y/2)={\rm
const}\times\tan(x/2)$, where we define this constant to be
$\tan\xi$:
\begin{align}
   \tan\xi = \tan(y/2)\cot(x/2).
   \label{xicoord.eqn}
\end{align}

  Equations (\ref{lambdacoord.eqn}) and (\ref{xicoord.eqn})
imply that
\begin{align}
   dx dy = J d\lambda d\xi,~
   J = \frac{1}{\sqrt{1-\beta\cos^2(2\xi)}},
   \label{Jacobian.eqn}
\end{align}
with $ \beta=1-\lambda^2/4$. As a cross check for this formula,
one can calculate the area of the BZ ($\int dxdy=4\pi^2$) in the
new coordinate system. A straightforward numerical integration
reassures us that the above Jacobian is correct and the intervals
$0<\xi<2\pi,~-2<\lambda<2$ count the original BZ only once.

Also the inverse transformation is given by
\begin{align}
  \cos x &= \frac{\lambda}{2}+\frac{J^{-1}-1}{\cos(2\xi)} ,\\
  \cos y &= \frac{\lambda}{2}-\frac{J^{-1}-1}{\cos(2\xi)} ,
\end{align}
where the Jacobian $J$ is already defined in equation
(\ref{Jacobian.eqn}). The variable $\xi$ can be interpreted as an
angle. In fact near the $\Gamma $ point where $x,y\approx 0$ and
hence, $\tan x\sim x,~\tan y\sim y$, we can actually see that
$\tan\xi\sim y/x$ and hence near to $\Gamma$ point $\xi$ can be
identified with the polar coordinate $\phi$. In this limit also we
can write $\lambda \approx 2-(x^2+y^2)/2=2-r^2/2$. Therefore our
coordinate system is a natural extension of polar coordinates
$(r,\phi)$. Near the zone center the energy contours become
circular, but near the Fermi energy $\lambda=0$, the contours are
rectangular.

  As an example of the application of this coordinate system,
let us calculate the exact DOS for SDW systems in 2D. All we have
to do is to switch to the new coordinate system so that
\begin{small}
\begin{align}
   \rho^{\rm 2D}(\nu)\! =\!\int\!\!\!\! \int\!\! \frac{dxdy}{4\pi^2}
   \delta\left(\varepsilon-\nu\right)
   \!=\!\int_0^{2\pi}\!\!\!\! d\xi\int_0^2\!\!\!d\lambda
   \frac{\delta\left(\varepsilon-\nu\right)}
   {4\pi^2\sqrt{1-\beta\cos^2(2\xi)}}.
\end{align}
\end{small}
The $\lambda$ integration can be performed by changing to the new
variable $\varepsilon=\sqrt{\Delta^2+\lambda^2}$ that gives
$d\lambda=\varepsilon d\varepsilon/\lambda$. Hence the final
result is
\begin{align}
   \rho^{\rm 2D}(\nu) =\frac{\nu}{\pi^2\sqrt{\nu^2-\Delta^2}}
   K\left(\sqrt{1-\frac{\nu^2-\Delta^2}{4}} \right),
\end{align}
where $K$ is the elliptic integral of first kind\cite{Abramowitz},
and the restriction $0<\lambda< 2$ translates to $\Delta<\nu<w_2$,
with $w_2^2=4+\Delta^2$ defining the upper band edge. This result
in the limit of $\Delta\to 0$ reduces to the appropriate result
for the tight binding bands\cite{Economou}. In this limit we have
a $\log(\nu)$ singularity at the Fermi surface. But for $\Delta\ne
0$, the nature of singularity near the lower band edge for 2D SDW
system is $\log (\nu-\Delta)/\sqrt{\nu-\Delta}$. In 1D, the
logarithmic contribution is absent. Hence the lower band edge in
2D spin density wave systems offers more DOS than 1D case. We see
in the text that this simple observation has profound implications
on third order nonlinearity in SDW systems.

  Calculation of DOS in 1D is much easier\cite{Economou}.
We need to calculate the following:
\begin{align}
   \rho(\omega) &=& \sum_k \delta(\varepsilon_k-\omega)
   = \int d\epsilon \rho_0(\epsilon)\delta(\varepsilon-\omega).
\end{align}
The DOS corresponding to gapless situation $\Delta=0$, is given by
$\rho_0(\omega)=1/\sqrt{1-\omega^2}$. Now we use the relation
$\epsilon=\sqrt{\varepsilon^2-\Delta^2}$ between energies
$\epsilon$ and $\varepsilon$ corresponding to $\Delta=0$ and
$\Delta\ne 0$ situations, respectively. This change of variables
gives:
\begin{align}
   \rho(\omega) = \int\!\! \frac{\varepsilon d\varepsilon}{\sqrt{\varepsilon^2-\Delta^2}}
   \frac{\delta(\varepsilon-\omega)}{\sqrt{w_1^2-\varepsilon^2}}
   =\frac{|\omega|}{\sqrt{\omega^2-\Delta^2}}
   \frac{1}{\sqrt{w_1^2-\omega^2}}\label{rho1D-Appendix.eqn},
\end{align}
where $w_1=\sqrt{1+\Delta^2}$. For SDW insulator, the gap
parameter $\Delta=-Um$ is determined by Coulomb correlation.

\end{document}